\documentclass{skaox2006}
\usepackage{graphicx}
\usepackage[official]{eurosym}
\newcommand{\ltsimeq}{\raisebox{-0.6ex}{$\,\stackrel
        {\raisebox{-.2ex}{$\textstyle <$}}{\sim}\,$}}

\begin{document}
   \title{The Square Kilometre Array}

   \author{Steve Rawlings\inst{1}
          \and
          Richard Schilizzi\inst{2}
          }

   \institute{Astrophysics, University of Oxford, Denys Wilkinson Building, Keble Road, Oxford, OX1 3RH, UK
         \and
             SKA Program Development Office, The University of Manchester, Manchester, M13 9PL, UK
             }

   \abstract{ 
We review the current status of the Square Kilometre Array (SKA) by outlining the science drivers for its Phase-1 (SKA$_{1}$) and 
setting out the timeline for the key decisions and milestones on the way to the planned start of its construction in 2016. We explain how Phase-2 SKA
(SKA$_{2}$) will transform the research scope of the SKA infrastructure, placing it amongst the great astronomical 
observatories and survey instruments of the future, and opening up new areas of discovery, many beyond the confines of conventional astronomy. 
}
   \maketitle
%
%
\section{Introduction and SKA status}
\label{sec:intro}

Over the last few decades, pulsar astronomy and Cosmic Microwave Background (CMB) observations have shown how radio astronomy can deliver profound results. Neutral Hydrogen (HI), pulsar and other observations with the Square Kilometre Array (SKA) have an exceptionally strong chance of continuing this story of scientific success. 

The quest for a radio telescope capable of imaging neutral Hydrogen (HI) to the limits of the observable Universe has been pursued for two decades (Wilkinson \cite{Wilkinson1991}), and astronomers appear to be at, or very close to, the point of  making statistical detections of HI at high redshift (Chang et al.\  \cite{Chang2010}; Paciga et al.\ \cite{Paciga2010}). This quest has been organized via a global Square Kilometre Array (SKA) program for which a science case much broader than HI and pulsars has been identified, refined and published (Taylor \& Braun \cite{Taylor1999}; Carilli \& Rawlings \cite{Carilli2004}). To obtain both the temperature sensitivity needed for HI and to find and time pulsars requires a huge (kilometre scale) essentially-fully-filled aperture, and thus requires that the SKA is built as a giant ground-based facility\footnotemark . 
\footnotetext{
For a filled-aperture telescope, the temperature, or surface brightness, sensitivity (measured in K) is independent of telescope size, explaining how 
fundamental breakthroughs in CMB astronomy have been made by modest-sized (few-m-class), millimetre-wavelength space telescopes like WMAP and Planck. The $\theta \sim \lambda / D$ law of diffraction means that studying similar angular scale (arcmin to degree) temperature fluctuations in redshifted HI (at metre wavelengths) requires telescopes of high filling factor ($\mu_{\rm f}$, the fraction of the physical area covered by the effective collecting area) and a minimum characteristic size of around a kilometre. Pulsar astronomy requires a similar centrally-concentrated collecting area to avoid computational-overload: for
pulsar searching the field of view (FOV) of each independent receptor must be covered by the smallest number of beams formed by adding or correlating
signals from the individual receptors, and this too is best achieved by a high-$\mu_{\rm f}$ kilometre-sized core.} 

At the time of the Crete meeting, the SKA effort was coordinated by the SKA Science and Engineering Committee, the Agencies SKA Group and an SKA Program Development Office funded by the international partners and an FP7 EC grant PrepSKA\footnotemark . 
\footnotetext{http://www.jb.man.ac.uk/prepska/} At the time of writing, the outputs of the policy work packages of PrepSKA - reports, including legal advice, concerning governance, procurement and funding - are currently being used to ensure that, before the end of PrepSKA in March 2012, full control of the SKA project will be passed to a new legal entity based at the SKA Project Office (SPO) to be housed in a new building at the University of Manchester's Jodrell Bank Observatory (JBO)\footnotemark . 
\footnotetext{http://www.skatelescope.org/}

The history of European funding for the SKA reflects its inclusion in the roadmap of the European Strategy Forum for
Research Infrastructures, and its ranking\footnotemark ~alongside the E-ELT as equal top priority for future large-scale
ground-based astronomy projects; 
\footnotetext{By the ASTRONET process, http://www.astronet-eu.org/} additionally
the US Astro2010 Decadal Survey noted that the SKA represents the  `long-term future for radio astronomy'. Elsewhere in the world, the SKA has excited considerable government-level interest, most notably in the two countries short-listed as potential hosts of the inner regions of the SKA: `Pathway to SKA' is the top priority for radio astronomy in
Australia, and significant funds have flowed into new initiatives such as the International Centre for Radio Astronomy Research (ICRAR\footnotemark), \footnotetext{http://www.icrar.org/} and the Australian SKA Pathfinder (ASKAP); the Heads of State of the African Union have acknowledged the importance of the SKA in the development of knowledge-based economies and South-African funds have flowed into an SKA-related
human capital development programme
and the South African SKA precursor MeerKAT.\footnotemark 
\footnotetext{http://www.ska.ac.za/} 

The schedule for the SKA features the following key decision points and milestones:

\begin{itemize}
\item April 2011. Having passed external international reviews - a System Level Conceptual Design Review (CoDR, Feb 2011) and a review of the pre-construction Project Execution Plan (the SKA PEP) - the SKA Founding Board was established to manage the transition from the current governance structure
to a new SKA legal entity.

\item Apr 2012. The SKA legal entity, operating from the SPO at the JBO, will decide on the site of the SKA.

\item Apr 2014. Construction funding to be approved for Phase 1 of the SKA: SKA$_{1}$, at a cost of 350 M\euro ~ [2007 purchasing power].

\item Apr 2016. SKA$_{1}$ construction begins.

\item Apr 2017. Construction funding to be approved for Phase 2 of the SKA: SKA$_{2}$, at an additional cost of 1.2 B\euro ~ [2007 purchasing power].

\item Apr 2018. SKA$_{2}$ construction begins.

\item 2020 onwards: full science operations with SKA$_{1}$.

\item 2024 onwards: full science operations with SKA$_{2}$.
\end{itemize}

The construction and operation of the SKA in two phases allows the science return to benefit optimally from the continued
gains in ICT capability with time, as is loosely encapsulated by `Moore's Law', and also allows the participation of the widest possible
collaboration of international funders, taking into account the spread in their likely profiles of spend on SKA. The SKA PEP includes an
Advanced Instrumentation Program (AIP) that allows for the further development and verification of  Aperture Array (AA) and smart-dish-feed technologies.

In brief summary (see also Garrett et al.\ \cite{Garrett2010} and Dewdney et al.\  \cite{Dewdney2010}) SKA$_{1}$ will consist of

\begin{itemize}

\item 50 AA stations (operating $\sim 70-450$ MHz) with a total physical collecting area $\sim 10^{6} ~ \rm m^{2}$, 50 per cent within a core of diameter $D_{\rm core} = 1 ~ \rm km$, with AA stations reaching out to a radius $r = 100$ km.

\item $250$ 15-m dishes (operating $\sim 0.3-10$ GHz), 50  per cent within $D_{\rm core} = 1 ~ \rm km$ and reaching $r=100$ km, potentially
(depending on the AIP)  with smart feeds allowing field-of-view or bandwidth expansion.

\end{itemize}

We emphasise in Sec.\ref{sec:science} that the 
science drivers of SKA$_{1}$ and SKA$_{2}$ remain those identified in Carilli \& Rawlings (\cite{Carilli2004}), but that this phased approach allows
some of the key science results to emerge once SKA$_{1}$ begins science operation towards the end of this decade.

Although we caution that temperature (rather than flux) sensitivity, or mapping speed, or other metrics are sometimes the correct figure of merit for SKA science,   
SKA$_{2}$ is defined by a requirement that across the full radio frequency range from 70 MHz to 10 GHz (with goals to extend to lower and higher
frequencies where technically feasible and affordable)
it has an r.m.s flux sensitivity $S_{\rm rms}$ (combining two polarizations for a point source) of:

\begin{equation}
S_{\rm rms} = \frac{100 ~ \rm mJy}{f  \sqrt{\delta \nu \delta t} },
\label{eq:one}
\end{equation}

\noindent
where $\delta \nu$ is the bandwidth (in Hz) and $\delta t$ is the exposure time (in s), and, by definition, $f=1$ for an  
SKA$_{2}$ realization with sensitivity $A_{\rm eff} / T_{\rm sys} \sim 20,000 ~ \rm m^{2} K^{-1}$ (as assumed within Carilli \& Rawlings \cite{Carilli2004}); the AA part of 
SKA$_{1}$ has $f \sim 0.1$ (assuming, away from the Galactic Plane, $T_{\rm sys} = 500$ K) and filling factor $\mu_{\rm f} \approx 0.8$, and the dish part of SKA$_{1}$ has $f \approx 0.05$. The working realization of SKA$_{2}$ (adapted from Schilizzi et al.\ 2007) has $f  \approx  [0.5({\rm AA}),0.5({\rm AA})+0.5({\rm dish}),0.5({\rm dish})]$ at frequencies $\nu = [0.13, 0.47 ,1.7]$ GHz and consists of the following: 250 low-frequency (70-450 MHz) AA stations (at least 125 in a $D_{\rm core}= 5$ km core); 250 mid-frequency (0.3-1.4 GHz) AA stations (at least 125 in a separate $D_{\rm core}= 5$ km core); and $2500$ 15-m dishes.

\section{Science Drivers: SKA$_1$ and SKA$_2$}
\label{sec:science}

Although the SKA science case is exceptionally broad (Carilli  \& Rawlings \cite{Carilli2004}), the design considerations of the SKA are determined by 
science drivers that exploit the unique opportunities radio astronomy brings to mapping out the origins of structure, to studying the physical laws and contents of the Universe, and to the discovery of new phenomena. 

The `21-cm line' of neutral Hydrogen (HI) is a cornerstone of the SKA$_{1}$ science case as its observation at high redshift allows 
astronomers to bridge a gap in studies of the distant Universe. This gap lies between $\sim 0.4$ Myr ($z \sim 1100$) after the Big Bang - when observations of the CMB show that the seeds of all structure were still tiny ($\sim 1$ part in $10^{5}$) perturbations in an otherwise smooth, featureless Universe - and the relatively recent Universe ($\sim 0.7-13.7$ Gyr after the Big Bang, or $z \sim 6-0$) in which observations show the Universe to be rich in galaxies, stars and planets. This gap is called the `Dark Ages' when the first stars, galaxies and black holes formed, and  
ends at $z \sim 10$ with the `Epoch of Reionization' (EoR); conditions in the EoR are probed indirectly by CMB polarization studies that deliver predictions regarding the evolving HI signal level. Directly mapping out the EoR is thus one of the two main science drivers for SKA$_{1}$ using the HI line redshifted to $\sim 100-200$ MHz (HI at $z \sim 13-6$; Carilli et al. \cite{Carilli2004a}).

How HI evolves outside this $13-6$ redshift range also constitutes basic information about the Universe. At $z>13$, observations become difficult because of the sky temperature and foregrounds, but also because, at these early times, the predicted HI signal level is increasingly speculative.
However, statistical information 
could plausibly be recoverable with SKA$_{1}$ out to $z \sim 19$ assuming a 
70 MHz lower-frequency cut-off (Garrett et al \cite{Garrett2010}). 

HI evolution between $z \sim 6-0$ can also be followed by SKA$_{1}$ in the 0.2-1.4 GHz waveband using both dishes and, potentially, the AAs. At these epochs the IGM is reionized, so the HI is likely be associated with galaxies, allowing straightforward utilisation of the `stacking technique', and incorporation of information from HI absorption 
to study the cosmic evolution of the ratio of atomic to molecular material involving sub-millimetre facilities like ALMA.
If SKA$_{1}$ were able to operate to above $\sim 15$ GHz\footnotemark 
\footnotetext{
The SKA design process currently has $10 ~ \rm GHz$ as the required upper frequency limit. However, we note that SKA dishes 
must also be capable of high-dynamic-range imaging, and thus have a target r.m.s surface accuracy of 0.5 mm (Garrett et al.\ \cite{Garrett2010}).
The Ruze-equation (Ruze \cite{Ruze1952}) for dishes indicates that their efficiency at 15 GHz will be 95 per cent of their efficiency value below 10 GHz.
}
then it is likely to also become the premier instrument for studying molecular material, through redshifted CO(1-0), in the EoR (Heywood et al.\  \cite{Heywood2011a} \cite{Heywood2011b}), and via cross-correlation techniques  
(Rawlings \cite{Rawlings2011}) this could contribute to the first detailed studies of the EoR. These would complement wider SKA studies of origins such as how galaxies evolved (Rawlings et al.\ \cite{Rawlings2004}), where cosmic magnetism originated (Gaensler et al.\ \cite{Gaensler2004}) and how planets formed in protoplanetary disks (Lazio et al.\ \cite{Lazio2004}).

The other cornerstone of the SKA$_{1}$ science case is pulsar astronomy\footnotemark . \footnotetext{
The technical demands placed on SKA$_{1}$ for pulsar astronomy -- most critically high sensitivity and time resolution, and requirement for correcting for effects such as dispersion and intrinsic time-varying properties of pulsars -- have strong overlap with the requirements of finding and studying other fast transient radio sources and hence extreme physics (Karastergiou et al.\ \cite{Karastergiou2011}). This means that the discovery of new transient phenomena, and the role of serendipity in such discoveries, can largely be pursued using an SKA designed around the requirements of pulsar astronomy (Cordes \cite{Cordes2011}; Kramer \cite{Kramer2011}).} 
Providing a facility capable of accurately timing pulsars - especially those in the Southern Hemisphere where the Galactic Centre is
easily observable, and where the SKA is the only planned high-sensitivity radio facility - is the second main science driver for SKA$_{1}$.
Through the discovery and follow-up timing of pulsars, the SKA is capable of making unique probes of fundamental physics (Kramer \cite{Kramer2011}; Karastergiou et al.\  \cite{Karastergiou2011}). The discovery of yet-more-exotic systems than the double-pulsar system (Lyne et al. \cite{Lyne2004}) - and specifically the discovery of any pulsar in the `strong-field limit' around a supermassive black hole - would, with SKA follow-up timing, push tests of Einstein's General Relativity into regimes where uncertain (e.g. Quantum Gravity) or completely new physics may become evident (Kramer et al.\  \cite{Kramer2004}). The SKA can expect to be centrally involved in the first detections of gravitational waves, an exceptionally promising target being the cosmic 
background due to the merging of supermassive black holes (see also Martinez-Sansigre \& Rawlings \cite{Alejo2011}), and ultimately to be able to pinpoint gravitational-wave sources (Sesana \& Vecchio \cite{Sesana2010}). 

The SKA$_{2}$ contribution to the study of fundamental laws can also utilise HI redshift surveys, containing a billion or so individually detected galaxies (Abdalla et al.\ \cite{Abdalla2010}) or statistical detection, via HI, of large-scale structure (Chang et al.\ \cite{Chang2010}) - to study key topics in modern physics. These include the nature of dark energy or post-Einstein gravity, neutrino mass, and the processes driving cosmic inflation (Rawlings et al.\  \cite{Rawlings2004}). 

It is a happy coincidence that HI and pulsars, as tools for unique studies the Universe, require that the SKA project constructs about one million square metres of collecting area, operating as an interferometer at frequencies from $\sim$70 MHz (for HI work) to at  least $\sim$10 GHz (for pulsar timing work). There must be two detector technologies: AAs at the lower frequencies; and dishes, potentially with smart 
(providing increased instantaneous bandwidth or field of view) feeds, at the higher frequencies.  Each detector technology needs one, and in the case of AAs probably two, compact (5-km diameter) cores containing about 50 per cent of the overall collecting area. This provides the raw temperature and flux sensitivity that is sufficient and necessary for the SKA to be able to: image HI structures in the 
EoR (Carilli et al.\ \cite{Carilli2004a}); study the evolution of HI across cosmic time; probe large-scale structure traced by HI in galaxies with sufficient accuracy
to constrain $w$ to less than 1 per cent accuracy (Abdalla et al.\ \cite{Abdalla2010}) and measure neutrino masses (Abdalla \& Rawlings \cite{Abdalla2007}); to find all the pulsars  in our galaxy (Kramer et al.\  \cite{Kramer2004}); 
to time the most exotic pulsar systems accurately enough to make fundamental new test of GR (Kramer et al.\  \cite{Kramer2004}); and to construct a 
Pulsar Timing Array (PTA)\footnotemark .\footnotetext{
Measurement of gravitational waves requires pulsar timing in both the Northern and Southern hemispheres to look for the large-scale angular  coherence
imprinted on pulsar timing residuals by low-frequency gravitational waves impinging on the Milky Way (Kramer et al.\ \cite{Kramer2004}). SKA$_{2}$ will provide world-leading pulsar-timing capability: enhancements to the existing Northern PTA, e.g. with FAST (Smits et al.\ \cite{Smits2009a}), providing complementary capabilities in the North.}
 
Thus, through both HI and pulsar astronomy, 
the SKA promises to open up areas of `post-Einstein science', and through probing regimes too extreme in size or energy scale 
to ever study on Earth, complements other fundamental science experiments such as those undertaken by the Large Hadron Collider at CERN.  With this 
ambitious vision, the SKA will deliver far more than the sum of  two or more distinct parts: e.g. a low-frequency ($\ltsimeq 1$ GHz) part for HI and a high-frequency part for pulsar work. The added value of a single system will already be evident in SKA$_{1}$: HI evolution studies above 450 MHz ($z \ltsimeq 2.1$) will use dishes not AAs, whilst surveys for Southern pulsars are likely to be performed by AAs as well as dishes [Smits et al.\ (\cite{Smits2009b}) assuming, following 
Garrett et al (\cite{Garrett2010}), an upper frequency of 450 MHz for the AA part of SKA$_{1}$].

The distribution of collecting area outside the SKA cores will be designed to  provide the resolution needed to ensure the main science aims are not compromised
by lack of spatial resolution or astrometric inaccuracy. This requires a range of longer baselines outside those available in the compact core. For the HI work, it will be necessary to separate HI in
in the intergalactic medium from HI in galaxies so as to observe the transition from distributed to confined-to-galaxy HI as Universal re-ionization progresses, requiring $\sim100$-km ($\sim 1$ arcsec resolution for AAs; $\sim 0.1$ arcsec resolution for dishes) baselines. For the pulsar work, astrometry is key to extracting the tightest constraints on GR from the timing follow-up, and so some very long  ($>1000$-km) baselines will be needed. 

The SKA will naturally become a flexible facility: capable of routinely imaging at `HST-like' ($\sim 0.1$ arcsec) resolution at higher frequencies; capable of non-confusion-limited resolution for surveys at low frequency where, for example, continuum surveys can detect all the quasars, radio-quiet or not, across most
 of the observable Universe (Jarvis \& Rawlings \cite{Jarvis2004}); and capable of participating in VLBI for astrometry
 and imaging with superb angular resolution. The SKA will naturally take its place alongside other great 
 upcoming observatories (e.g. ALMA, JWST and ELTs) and the other future premier astronomy survey instruments (e.g. Euclid, LSST).

\section{Towards SKA: SKA$_{0}$}
\label{sec:ska0}

There are a large number of technology demonstration programs and telescopes at various stages of completion that lie outside the direct PrepSKA `Description of Work' but which have been identified as pathfinder activity for the SKA. In the case of telescopes under construction on the potential host sites, these are termed precursor facilities. It is a convenient simplification to refer to these collectively as SKA$_{0}$ since they are all doing 
SKA-relevant work under independent funding and direction.

\begin{itemize}

\item ASKAP: Australian dish precursor consisting of 36 12-m dishes ($f \sim 0.005$) . Key technical pathfinding: inexpensive high-performance dishes; phased array feeds (PAFs) and associated digital back-end and data transport solutions. Key science pathfinding: large sky-area HI and continuum surveys.

\item MeerKAT: South-African dish precursor consisting of 64 13.5-m dishes ($f \sim 0.01$). Key technical pathfinding: inexpensive high-performance dishes; single-pixel feeds and associated digital back-end and data transport solutions. Key science pathfinding: deep HI and continuum surveys; pulsar timing.

\item WSRT/APERTIF: PAF system (1-1.75 GHz) to be installed on 12 of the 14 WSRT antennae ($f \sim 0.007$) to increase its survey speed by a factor 20. Key technical pathfinding: PAFs and digital backends. Key science pathfinding: wide-field Northern HI and continuum surveys.

\item MWA: USA and Australian AA precursor (80-300 MHz with one antenna type) to consist of 512 tiles (each with 16 antennae, so $f \sim 0.001$
and filling factor $\mu_{\rm f} \sim  0.01$ within 1-km) on candidate Australian SKA site. Key technical pathfinding: small amounts of signal aggregation prior to  correlation-rich processing. Key science pathfinding: EoR experiment.

\item LOFAR: Dutch-led, but European, AA pathfinder (15-240 MHz with two antenna types) with a core in the northern Netherlands and stations in several European Countries ($f \sim 0.01$ and $\mu_{\rm f} \sim  0.02$ within 1-km).  Key technical pathfinding: beam-forming-rich signal processing. Key science pathfinding: EoR experiment; wide-area continuum surveys.

\item e-MERLIN: UK-based dish array consisting of at least 6 30-m-class dishes (plus the 76-m Lovell Telescope, in total $f \sim 0.01$)
connected in real time to a JBO-based correlator via a dark fibre system. Key technical pathfinding: data and timing-signal transport via dark fibres. Key science pathfinding: high-resolution, wide-field imaging with dishes.

\item e-EVN: European-dominated, but increasingly global, system connecting 30-m and larger class dishes in real time to a correlator at the Joint Institute for
VLBI in Europe (JIVE) in the northern Netherlands ($f  \sim 0.05$).  Key technical pathfinding: data and timing-signal transport via internet. Key science pathfinding: ultra-high-resolution, wide-field imaging with dishes.

\item European Pulsar Timing Array: Pan-European collaboration involving the five largest radio telescopes in Europe ($f \sim 0.03$): Effelsberg, WSRT, Nancay, Lovell, Sardinia. Key technical pathfinding: combining large dishes in a PTA. Key science pathfinding: timing and astrometry of Northern pulsars.

\item FAST: Giant Chinese dish telescope (0.3-5 GHz) capable of covering sky up to 40 degrees from the Zenith with $f \sim 0.1$.
Key technical pathfinding: large dishes on long baselines. Key science pathfinding: timing and astrometry of Northern pulsars.

\end{itemize}

\section{Concluding Remarks}
\label{sec:conc}

In this paper we have motivated the specifications for the SKA, and its phased realization as Phase-1 SKA (SKA$_{1}$) and
Phase-2 SKA (SKA$_{2}$), by the transformational science we know it can deliver. The difference between this coherent approach and the diverse
range of pathfinding activity summarised in Sec. \ref{sec:ska0} is not accidental, indeed we believe it is a critical step towards making the SKA happen
{\em\bf now} rather than at an indeterminate point in the future. The recent formation of the SKA Founding Board signals the seriousness of this 
endeavour.

SKA$_{1}$ will allow HI studies to map out the Epoch of Re-ionization (EoR) and open up the Dark Ages of the Universe; it will also use pulsars to make unique tests of GR, including an extremely 
good chance of being critical to the first detections of gravitational waves using PTAs. As a bonus there will be a wealth of other astronomy possible with SKA${_1}$ including
all-Southern-sky arcsec-resolution radio continuum surveys identifying counterparts to the next-generation of photometric (e.g. LSST) and spectroscopic surveys planned in other astronomical wavebands, and sub-arcsec-resolution imaging capabilities needed to complement other great observatories like ALMA, ELTs and the JWST. We illustrate these synergies with just one example: ALMA and SKA are `natural integral-field-unit (IFU) spectrometers'
providing data cubes showing the location and kinematics of molecular, atomic and ionized material in distant galaxies: early-light E-ELT IFU instruments like HARMONI (Thatte \cite{Thatte2011}) will image individual HII regions and deliver kinematics from stellar and gas tracers -- all these facilities will be
needed to get a comprehensive picture of galaxy formation and evolution.

The plan for SKA$_{1}$ ensures breakthrough science early in the project, but the truly transformational science return will need the full capabilities of SKA$_{2}$. SKA$_{2}$ will probe the Dark Ages from $z=6$ to perhaps $z \sim 20-30$. It will use HI to make the `billion galaxy'  surveys needed to address key questions such as neutrino mass and sub-per-cent accuracy on the dark energy $w$ parameter. It will make deep polarisation-sensitive `all-southern-sky' surveys that will probe the origin of magnetic field in the Universe, determining its role in the formation of structure from the cosmic web, through galaxies, to stars. For pulsar surveys, SKA$_{2}$ will deliver the full census of $\sim 20,000$ normal and $\sim 1000$ millisecond pulsars in our Galaxy, with a concomitant increase in chances of finding the rare `holy grail' systems that allow precision tests of GR and measurement of the equation of state of nuclear matter at supranuclear densities. Pulsar timing will be revolutionised by SKA$_{2}$ capabilities allowing fundamental studies of the 
graviton and the ability to pinpoint gravitational wave sources. SKA$_{2}$ will also be a critical instrument for understanding planet formation and astrobiology. 

There is a strong expectation that, as the SKA is realised in phases, that the community using it will expand. Space scientists will
use the SKA to track spacecraft and near-earth objects and to monitor space weather, embedding SKA in Solar-System
exploration, and space safety and security systems (Butler et al.\ \cite{Butler2004}; Jones \cite{Jones2004}). The SKA's visibility as a research infrastructure for fundamental physics will develop not only through its key science topics but in other ways such as the plans to monitor radio flashes from the moon that are expected to 
result from interactions between ultra-high-energy neutrinos and the Moon's regolith (Falcke et al.\  \cite{Falcke2004}). 

The societal impact of SKA research will be maximised 
by fostering a close relationship between SKA and `Citizen Science': by expanding on initiatives such as `SETI@home'\footnotemark
\footnotetext{http://setiathome.berkeley.edu/}
and `GalaxyZoo'\footnotemark, the SKA can expect 10's of millions of individuals in global society to get truly involved by donating their CPU cycles and brain power to SKA data re-processing and science.
\footnotetext{http://zoo1.galaxyzoo.org/}

With the SKA, as with other research infrastructures of the future, 
there is an expectation that its design, construction and development will be implemented by multi-sectoral consortia 
of industry and academic institutions carrying out packages of work under the direction and authority of the SPO. By optimising the
involvement of industry and academia across the World, this should help ensure that the SKA
has the widest possible impact in the form of technology spin-off and human capital development.
As well as revealing profound truths about the Universe, we hope the SKA project will generate new ways of doing ICT, using 
green energy wherever feasible.

\begin{acknowledgements}
The authors thank the EC for supporting the SKA project via the FP7 PrepSKA project. 
\end{acknowledgements}

\end{document}